\documentclass [12pt] {report}
\usepackage{amssymb}
\newcommand {\fr} {\displaystyle\frac}

\newcommand {\be} {\begin{equation}}
\newcommand {\ee} {\end{equation}}
\newcommand {\ba} {\begin{array}}
\newcommand {\ea} {\end{array}}
\newcommand {\bp} {\begin{picture}}
\newcommand {\ep} {\end{picture}}
\newcommand {\bc} {\begin{center}}
\newcommand {\ec} {\end{center}}
\newcommand {\bt} {\begin{tabular}}
\newcommand {\et} {\end{tabular}}
\newcommand {\lf} {\left}
\newcommand {\rg} {\right}

\newcommand {\bls} {\baselineskip}

\newcommand {\cI} {{\cal I}}
\newcommand {\cM} {{\cal M}}

\newcommand {\cL} {{\cal L}}
\renewcommand{\th}{\theta}
\newcommand {\ses} {\medskip}

\newcommand {\bibit} {\bibitem}
\newcommand {\nin} {\noindent}

\def\2#1#2#3{{#1}_{#2}\hspace{0pt}^{#3}}
\def\3#1#2#3#4{{#1}_{#2}\hspace{0pt}^{#3}\hspace{0pt}_{#4}}
\newcounter{sctn}

\begin {document}

\begin {titlepage}

\vspace{0.1in}

\begin{center}
{
FINSLERIAN  EXTENSION OF  LORENTZ TRANSFORMATIONS
}\\
\end{center}

\begin{center}
{
AND FIRST-ORDER CENSORSHIP THEOREM
}\\
\end{center}

\vspace{0.3in}

\begin{center}

\vspace{.15in}
{\large G.S. Asanov\\}
\vspace{.25in}
{\it Division of Theoretical Physics, Moscow State University\\
117234 Moscow, Russia\\}
\vspace{.05in}

\end{center}

\begin{abstract}
Granted the post-Lorentzian relativistic kinematic transformations are
described in the Finslerian framework,
the uniformity between the actual light velocity anisotropy change
and the anisotropic deformation of measuring rods
can be
the reason proper
for the null
results of the Michelson-Morley-type experiments
at the first-order level.
\ses

\nin
\ses

\end{abstract}

\end{titlepage}

\vskip 1cm

{\nin\bf 1. Introduction}
\medskip

\nin
One of the important goals of modern physics is to accurately investigate
and measure possible post-Lorentzian relativistic parameters [1,2].
In this respect, however, the outcomes of
the Michelson-Morley-type experiments
(including
various optical interferometer experiments [3-9] as well as the modern
laser high-precision experiments with improved results [10-13])
seems to be pessimistic rather than optimistic: experiments steadily reproduce
``no fringe shift" and, therefore,
the ``null result" for deviations from the traditional
theory of Special Relativity (SR).

If it is just the Finslerian extension of SR that turns out to be the case,
then it will perhaps be possible to infer
rigorously, and geometrically,
that the light velocity value
should be
anisotropic in moving reference
frames.
Would it be testable?

The important point is that
the experiments
don't measure directly the differences
between the  velocities of light signals sending in different directions.
Primarily, the experiments measure the fringe shifts corresponding to
 differences between the times of signal flights.
The values of the times are, however,
results of comparing between the light velocity values, on the one hand,
and the  lengths of interferometer legs (optic paths), on the other hand.
There is an open possibility that these two major factors can compensate
one another
(here the analogy with the Lorentz-Fitzerald contraction is pertinent; [14,15])
and lead, therefore, to the conclusion that no apparent fringe shifts
should come about.

Mixing the two aspects of anisotropy would, generally, lead to erroneous
identifications.
Alas, having assumed that,
though the isotropical
symmetry is implied for a preferred rest frame $S_0$,
the observation space
in a moving reference frame $S(v)$
may devoid of spatial isotropy and that, therefore,
the anisotropy of light velocity value
in $S(v)$ may well be expected,
the assumption that the length of a measuring rod
(as well as of a interferometer leg or a optic path)
should not be effected by rotation performed in $S(v)$
becomes neither obvious nor cogent.

On explaining relevant arguments and evaluation methods in Section 2,
we are able in
Section 3 to assign the theoretical realism to
\ses

\nin \bf THE RELATIVISTIC FIRST-ORDER CENSORSHIP THEOREM.
\it The Finslerian extension of Lorentz transformations
retains conspiracy of the first-order effects
in the Michelson-Morley-type experiments.\rm
\ses

We use the particular special-relativistic Finslerian metric function
$F_{SR}(g;T,X,Y,Z)$, where $g$ is a parameter,
subject to the following attractive conditions:
\ses

\nin
 (P1) {\it The indicatrix-surface $\cI$ defined by the equation
$F_{SR}(g;T,X,Y,Z)=1$
is a regular space of a
constant negative
curvature, $R_{\cI}$},
\ses

\nin
which is a convenient stipulation to seek for the nearest
 Riemannian-to-Finslerian relativistic
 generalization;
\ses

\nin
 (P2)
{\it The Finslerian metric function is compatible with the principle of spatial isotropy};
\ses

\nin
 (P3) {\it The associated Finslerian metric tensor is of the time-space signature $(+---)$};
\ses

\nin
 and
\ses

\nin
 (P4) {\it The principle of correspondence holds true},
\ses

\nin
 that is, the associated Finslerian metric tensor [16,17]
reduces exactly to its ordinary known special-
 or general-relativistic prototype when $R_{\cI}\to-1$, which physical
 significance is quite transparent.\ses

 All the items (P1)--(P4) are obeyed whenever one makes the choice
\be
F_{\rm SR}(g;T,X^1,X^2,X^3)=TV(g;|{\bf X}|/T)
\ee
  with
\be
V(g;w)=[Q(g;w)]^{1/2}\lf(\fr{1+g_-w}{1+g_+w}\rg)^{-g/4h}
 \equiv(1+g_-w)^{g_+/2h}(1+g_+w)^{-g_-/2h},
\ee
where
\be
Q= 1-gw-w^2\equiv (1 +g_-w)(1 +g_+w)
\ee
and
\be
g_\pm=-\fr12g\pm h, \quad h=\lf(1+\fr14g^2\rg)^{1/2}.
\ee

Vice versa, we can claim the following\ses
\ses

\nin
 {\bf THE UNIQUENESS THEOREM.}
 \it The properties \rm(P1)--(P4), \it when treated as conditions imposed on
 the Finslerian metric function, specify it unambiguously in the form
given by Eqs.
\rm(1.1)--(1.4),\\[2mm]
 \rm which is valid due to Theorem~5 of the work~[18] (see also~[19--21])
 in which  the Finsler spaces compatible
 with the properties (P1) and (P2) have been studied.
The indicatrix curvature is
 \be
 R_{\cI}=-\lf(1+\fr14g^2\rg)\le -1.
 \ee
\ses
\ses
\ses
\ses

\setcounter{sctn}{2}
\setcounter{equation}{0}
{\nin\bf 2. RELATIVISTIC KINEMATIC PATTERNS}
\ses

{\bf 2.1. The Lorentz transformations} have been
 serving over a century to
 predict and
describe
new relativistic effects. Despite the general feeling of a high
 degree of accuracy between predictions and measurements, various
 modifications, including the well-known cases
\ses

\nin
(I) The Robertson Transfomations [22];
\ses

\nin
(II) The Edward Transformations [23, 24];
\ses

\nin
(III) The Mansouri--Sexl Transformations [25];
\ses

\nin
(IV) The Tangherlini Transformations [26];
\ses

\nin
and
\ses

\nin
(V)  The Selleri Transformations [27]
\ses

\nin
(listed here in the chronological order) have been used to overcome
traditional Lorentzian patterns.
The
 transformations (I)--(III) have clearly been compared with each other in~[28];
 a~systematic review of various kinematics relations stemed from the choice
 (IV) can be found in~[29].
In fact, the transformations (I)--(V) have been
 introduced primarilly to reanalyze the role of synchronization procedure
(see [30-33]),
and examine possible observable difference
which  would result if the light speed were anisotropic.
The case (V) was examined systematically in [27] under attractive conditions:
namely, assuming that the two-way light velocity is invariant under
transformations
among inertial reference frames and that the clock relativistic retardation
takes place with the ordinary square-root factor,
$\sqrt{1-\beta^2}$,
the transformations (V) ensue.
No Finslerian
metric function
matching a member of the set of non-Lorentzian transformations (I)--(V) has
 been proposed.

However, the geometrically-motivated status of SR
can be retained
under post-Lorentzian
extension
 if the Finsler geometry is invoked as a necessary basis to proceed, which
leads to choose the case $F=F_{SR}$ as given by (1.1)--(1.4).
\ses

{\nin\bf 2.2. The Finslerian kinematic transformations}
\be
T=\fr{1}{V(g;v)}t+\fr1{V(g;v)}v x,
\qquad X=\fr1{V(g;v)}v t+\fr{1}{V(g;v)}(1-gv)x,
\ee
\bigskip
\be
Y=\fr{\sqrt{Q(g;v)}}{V(g;v)}y,
\qquad
Z=\fr{\sqrt{Q(g;v)}}{V(g;v)}z,
\ee
can well be used to accomplish
the special-relativistic transition from
 a preferred isotropical
rest frame $S_0$ to a reference frame $S(v)$
moving inertially at
a relative velocity $v>0$ with respect to $S_0$
along the common direction of the $x$- and $X$-axes.
The Capital letters, $\{T,X,Y,Z\}\in S_0$,
represent measuremenets in $S_0$, while the small letters,
$\{t,x,y,z\}\in S(v)$,
represent measurements in $S(v)$.

The transformations (2.1)-(2.2)
can unambiguously be explicated from the Finslerian metric function
$F_{SR}$
given by (1.1)-(1.4): to this end one should
merely calculate the tetrads of the associated Finslerian metric
tensor to use them for connecting the proper reference systems
of the RFs $S_0$ and~$S(v)$ (cf. the general relativistic role of tetrads,
[34-35]).

Inversion yields
\be
t=a(g;v)T+e(v)x, \quad x=b(g;v)(X-vT),\quad y=d(g;v)Y,\quad z=d(g;v)Z
\ee
with
\be
a=V(g;v),\quad b(g;v)=\fr{V(g;v)}{Q(g;v)},\quad d(g;v)=\fr{V(g;v)}{\sqrt{Q(g;v)}},
\ee
and
\be
e(v)=-v.
\ee
The kinematic meaning of the caracteristic Finslerian
parameter $g$ is the following:
\be
g=(db/dv)\big|_{v=0};
\ee
the normalizalion conditions
$
 a(0) = b(0) = V(0) = 1
$
have been implied; we have also put
$
(da/dv)\big|_{v=0}=0
$
to agree perfectly with the  low-velocity experimental evidence.

Whenever
$v\ll1$, we obtain
\be
a(g;v)=1-\fr12v^2-\fr13gv^3-\fr18(1+2g^2)v^4+O(5),
\ee
\ses
\be
 b(g;v)=1+gv+\fr12(1+2g^2)v^2+O(3),
\ee
\ses
\be
 d(g;v)=1+\fr12gv+\fr38g^2v^2+O(3),
\ee
where
the
symbol $O(N)$ in (2.7)--(2.9) denoted the terms proportional to $v^K$
with $K\ge N$.
 The neglect
of the third-order term in the low-velocity expansion of the time
dilatation factor $a$ would entail the traditional pseudo-Euclidean
case: $\{g = 0,\quad a(v) = \sqrt{1-v^2},\quad b(v) = 1/a(v)\}$.
\ses
\ses
\ses
\ses

\setcounter{sctn}{3}
\setcounter{equation}{0}
{\nin\bf 3. METROSURFACE AND LIGHTSURFACE}
\ses

In each RF $S(v)$ an observer
is assumed to be equipped with a Metrosurface, $\cM_v$,
to assign the length values to roads
(or legs, or optic paths,...) pointed in various directions. The Finslerian approach
proposes \it
a geometrically-motivated method \rm
 to introduce the device according
to the definition
\be
\fr1{V(g;v)}
F_{SR}\Bigl(g;vx,(1-gv)x,\sqrt{Q(g;v)}\,y,\sqrt{Q(g;v)}\,z\Bigr)=1.
\ee
Indeed, it is natural to consider $\cM_v$ to be
a geometric place of the ends of segments directed from the origin, $O$,
of $S(v)$ subject to the conditions that the transforms of the segments
from
$S(v)$ into $S_0$ have the unit Finslerian length:
$F_{SR}(g;T,X,Y,Z)=1.$
When we  use here the substitutions (2.1)-(2.2)
and put $t=0$, we just receive (3.1).

On using (3.1), simple calculations show that in the approximation
keeping only first powers of the parameter $g$,
the polar-angle representation for $\cM_v$ reads
\be
r(g;v;\th)
=
1+\fr12gv(1-\cos\th)^2,
\ee
where
$\th$ is the angle
made in $S(v)$
between the $x-$axis and the direction of the measuring segment
(rod, leg, optical way,...). In the pseudo-Euclidean limit, that is when $g=0$,
the Metrosurface is simply the unit sphere, $r=1$, in any $S(v)$.

Let us now pose the question of \it
what is the form of a similar representation
for the Lightsurface, $\cL_v$, in $S(v)$?
\rm
The due understanding of the question can be gained on
borrowing  the ordinary special- and generel-relativistic
treatment of the Lightsurface to be the isotropic surface of the
fundamental metric function, that is, in our present study,
the surface defined
by $F_{SR}=0$. Then from (1.2) it follows that the light velocity in $S_0$ is
the constant
\be
c=g_+
\ee
(because of $V(g;g_+)\equiv 0$). Transforming  the resultant equation
$
X^2+Y^2+Z^2=(g_+)^2T^2
$
with the help of (2.1)-(2.2))
yields the light
velocity value
$c(g;v;\th)$ in $S(v)$:
\be
\fr{c(g;v;\th)}{g_+}=
1+\fr12gv(1-\cos\th)^2.
\ee

Comparing between (3.2) and (3.4) reveals the uniformity of anisotropy of
the two types mentioned, such that
\be
\fr{r(g;v;\th)}
{c(g;v;\th)}=const
\ee
in any admissible reference frame $S(v)$, which thereby
proves the Censorship Theorem formulated above in Section 1.
\ses
\ses
\ses
\ses

{\nin\bf 4. CONCLUSIONS}
\ses

The Finslerian approach explicitly provides rigorous
definitions of the means whereby the
special-relativistic quantities can be measusred
and, therefore,  invites the re-consideration of the SR to substitute the
Lorentz transformations with the Finslerian kinematic transformations.

In the foregoing, we have
defined due devices, namely the Metrosurface and the Lightsurface,
built on the post-Lorentzian Finslerian transformation properties
of standars rods and light rays, to set forth the operational basis for
the standard relativistic measurements.
The light velocity anisotropy
and
the spatial length anisotropy
may occur to be different concepts in the RF~$S$, so that, when
analyzing some relativistic observations, one should bring to
evidence what kind of the anisotropy
is considered in the corresponding experimental set-up.

{\it Nature may reveal conspiracy of  anisotropy of space-time!}
The breakdown of spatial isotropy in a moving RF should, in principle,
contribute
to both the light-velocity anisotropy and to the length standard anisotropy.
In the context of the analysis and calculations
described above,
these two effects just compensate one another
in the first-order treatment, thereby making one obtains in any inertial
reference frame the null result
expected for spatial isotropy.
To settle the matter outright, it is required to conduct experiments
measuring the
second-order relativistic effects.
\ses
\ses
\ses
\ses
\ses
\ses

\def\bibit[#1]#2\par{\rm\noindent
                     \parbox[t]{.05\textwidth}{\mbox{}\hfill[#1]}\hfill
                     \parbox[t]{.925\textwidth}{\bls11pt#2}\par}

{\nin\bf REFERENCES}\\
\ses

\nin
 1. M.P. Haugan and C.M. Will: \it Physics Today \bf 40\rm, 69 (1987).\\
 2. C.M. Will: \it Phys. Rev. D \bf 45\rm, 403 (1992).                   \\
 3. A.A. Michelson and E.W. Morley,
\it Amer. J. Sci. \bf 34\rm, 333  (1887).\\
 4. B. Jaff\'e: \it Michelson and the Speed of Light \rm (Anchor Books,
Doubleday, N.\,Y., 1960).                                        \\
 5. R. J. Kennedy and E. M. Thorndike, \it Phys. Rev. B \bf 42\rm, 400 (1932).\\
 6. H. E. Ives and G. R. Stillwelt, \it J. Opt. Soc. Amer. \bf 28\rm,
215 (1938);
\bf 31\rm, 369 (1941).                                                          \\
7. E.W. Silvertooth, \it J. opt. Soc. Am. \bf 62\rm, 1330 (1972).    \\
8. S. Marinov, \it Czech. J. Phys. B \bf 24\rm, 965 (1974).\\
9. D.J. Mora, \it Not. T. Astr. Soc. \bf 173\rm, 33 (1975).  \\
10. T.S. Jaseja, A. Javan, J. Murray, and C.H. Townes:
\it Phys. Rev. A \bf 133  \rm  ( 5), 1221
(1964).\\
11. A. Brillet and J.L. Hall: \it Phys. Rev. Lett. \bf 42\rm, 549 (1979).\\
12. T.P. Krisher et al.: \it Phys. Rev. D \bf 42\rm, 731 (1990).            \\
 13. D. Hils and J.L. Hall: \it Phys. Rev. Lett. \bf 64\rm, 1697 (1990).     \\
 14. H.A. Lorentz: \it The Theory of Electrons and its Application to
the Phenomena of Light \rm
and Radiant Heat, 2nd edition (Dover, N.Y., 1952), p. 178. \\
 15. C. M{\o}ller: \it The Theory of Relativity\rm, 2nd edition
(Claredon, Oxford, 1972), p. 21. \\
 16. H. Rund: \it The Differential Geometry of Finsler
 Spaces \rm (Springer, Berlin, 1959). \\
 17. G.S. Asanov: \it Finsler Geometry, Relativity and Gauge
 Theories \rm (D.~Reidel Publ.
 Comp., Dordrecht, 1985).\\
  18. G.S. Asanov: \it Aeq Math. \bf49\rm, 234 (1995).\\
  19. G.S. Asanov: \it Rep. Math. Phys. \bf 39\rm, 69 (1997);
\bf 41\rm, 117 (1998);
\bf 46\rm, 383 (2000);
\bf 47\rm, 323 (2001).\\
  20. G.S. Asanov: \it Moscow University Physics Bulletin
 \bf49 \rm (1), 18 (1994); \bf 51 \rm (1), 15 (1996);
\bf 51 \rm (2), 6 (1996); \bf 51 \rm (3), 1 (1996);
\bf 53 \rm (1), 15 (1998).\\
21. G.S. Asanov: ``The Finsler-type recasting of Lorentz transformations."
In: Proceedings
 of Conference  {\it Physical Interpretation of
Relativity Theory}, September 15-20 (London, Sunderland, 2000), pp. 1-24.\\
 22. H.P. Robertson: \it Rev. Mod. Phys. \bf 21\rm, 378 (1949).\\
 23. W.F. Edwards: \it Am. J. Phys. \bf31\rm, 482 (1963).\\
  24. J.A. Winnie: \it Philos. Sci. \bf37\rm, 81 and 223 (1970).\\
  25. R. Mansouri and R. Sexl:
\it Gen. Rel. Grav. \bf8\rm, 496, 515, and 809 (1977) .\\
 26. F.R. Tangherlini: \it Suppl. Nuovo Cimento \bf20\rm, 1 (1961).\\
 27. F. Selleri: ``Space and time should be preferred to
spacetime, 1 and 2". In: {\it Redshift
and gravitation in a relativistic universe}
(K. Rudnicki, ed.) (Apeiron, Montreal, 2001),
pp. 63-71 and 81-94.\\
 28. Y.Z. Zhang: \it Gen. Rel. Grav. \bf27\rm, 475 (1995).\\
 29. G. Spavieri: \it Phys. Rev. A \bf 34\rm, 1708 (1986).   \\
30. H. Reichenbach: \it The Philosophy of space and time \rm (Dover Pupl.,
Inc., N.\,Y., 1958).                            \\
31. A. Gr\"unbaum: \it The Philosophy of Science, \rm
A.~Danto and S.~Morgenbesser,
eds. (Meridian Books, N.\,Y., 1960).\\
32. A. Gr\"unbaum: \it The Philosophy of Space and Time
\rm (Redei, Dordrecht, 1973).         \\
33. R. Anderson, I. Vetharaniam and G.E. Stedman: \it  Phys. Reports
\bf 295\rm, 93-180 (1998).             \\
 34. H.-J. Treder, H.-H. von Borzeszkowski, A. van der Merwe, and W~Yourgrau:
\it Funda-
mental Principles of General Relativity Theories. Local and Global
Aspects of Gravitation
and Cosmology \rm (Plenum, N.\,Y., 1980).\\
 35. J.L. Synge: \it Relativity: the General Theory
\rm (North-Holland, Amsterdam, 1960).

\end{document}